\documentclass[preprint]{aastex62}

\usepackage[utf8]{inputenc}
\usepackage[T2A]{fontenc}
\usepackage[english]{babel}

\usepackage{needspace}

\usepackage{url}
\usepackage{graphicx}
\usepackage{hyperref}
\usepackage{amsmath}
\usepackage{gensymb}
\usepackage{multirow}
\newcommand{\ellps}{\multicolumn{1}{c}{...}}

\usepackage[capitalise, nameinlink, noabbrev]{cleveref}
\crefname{equation}{eq.}{eq.}
\usepackage{booktabs}
\usepackage[flushleft]{threeparttable}

\usepackage{natbib}
\newcommand{\vecc}[1]{\mathbf{#1}}

\newcommand{\omegap}{\omega_{\mathrm p}}
\newcommand{\vgroup}{v_{\mathrm{gr}}}
\newcommand{\ngroup}{n_{\mathrm{gr}}}
\newcommand{\Ne}{N_{\mathrm{e}}}
\newcommand{\Ncrit}{N_{\mathrm{crit}}}
\newcommand{\Rsun}{R_\odot}
\newcommand{\dd}{\mathrm{d}}

\begin{document}

\title{
    Exploring the Asymmetry of the Solar Corona Electron Density \\
    with Very Long Baseline Interferometry
}

\correspondingauthor{Dan Aksim}
\email{danaksim@iaaras.ru}

\author{Dan Aksim}
\author{Alexey Melnikov}
\author{Dmitry Pavlov}
\author{Sergey Kurdubov}
\affiliation{Institute of Applied Astronomy, St. Petersburg, Russia}

\begin{abstract}

The Sun's corona has interested researchers for multiple reasons,
including the search for solution for the famous coronal heating problem
and a purely practical consideration of predicting geomagnetic storms on Earth.
There exist numerous different theories regarding the solar corona;
therefore, it is important to be able to perform comparative
analysis and validation of those theories.
One way that could help us move towards the answers to those problems
is the search for observational methods that could obtain information about
the physical properties of the solar corona and provide means for comparing
different solar corona models.

In this work we present evidence that VLBI observations are,
in certain conditions, sensitive to the electron density of the solar corona
and are able to distinguish between different electron density models,
which makes the technique of VLBI valuable for solar corona investigations.
Recent works on the subject used a symmetric power-law model of the electron
density in solar plasma; in this work, an improvement is proposed based on
a 3D numerical model.

\end{abstract}


\section*{\phantom{a}}
\vspace{-0.7cm}

\noindent\makebox[\textwidth][c]{%
\begin{minipage}{15.5cm}
\footnotesize{
\noindent
This is the version of the article accepted for publication including
all changes made as a result of the peer review process, and which may also
include the addition to the article by IOP Publishing of a~header,
an article ID, a cover sheet and/or an ‘Accepted Manuscript’ watermark,
but excluding any other editing, typesetting or other changes made by IOP
Publishing and/or its licensors.
}
\end{minipage}
}

\vspace{0.5cm}

\section{Introduction}

Since the discovery of the solar corona in the 19th century to this day, the
physical processes happening inside it are not completely understood.  In
particular \citep[see e.g.][]{Cranmer2002}, the cause of heat of the
$10^6$~K of the corona and acceleration of the solar wind is not known
with certainty. While many theories were proposed, and more continue to appear
and compete with each other, observational data is needed to eliminate wrong
assumptions. There are very useful and important observations of solar corona
by spacecraft, like STEREO \citep{Harrison2008}; moreover, \textit{in situ} measurements
of the solar corona by Parker Solar Probe \citep{Bale2016} are expected soon.
However, it is also possible to perform high-quality measurements of the solar
corona from Earth; that is, to measure one particular property of the
corona---the electron density---along a ray path.

The idea of determining electron densities in space plasmas
by means of measuring group time-delays of radio signals traveling
between a transmitter located on Earth and an interplanetary spacecraft
was formulated at the times of the first spacecraft launches \citep{Kelso1959}.
Since then, numerous works regarding the use of radio measurements
for determining the electron density in one particular space plasma,
the solar corona, have been published.
Those works were traditionally based on analysis of single- and dual-frequency
time-delay observations of spacecrafts orbiting Venus and Mars.
An overview of some of such works is given by \citet{Bird1990, Bird2012}.

Besides spacecraft tracking, a technique that measures time delays of radio
signals and could, in theory, determine the solar corona's electron density,
is very long baseline interferometry (VLBI).
The quantity measured by VLBI is the difference in arrival times of
a signal from an extragalactic radio source received at two radio telescopes
\citep{Schuh2012}.
As the spatial distribution of the electron density of the solar corona is clearly
not uniform, the dispersive delays accumulated by radio signals passing
through different regions of the corona would also be different,
and measuring these differences should allow for reconstruction
of the electron density.

Though yet not particularly common, the idea of utilizing VLBI
for solar corona electron density measurements has been adopted in several recent
works.
Specifically, the parameters of a simple power-law model for the electron
density were estimated using data of multiple VLBI sessions from 2011--2012
\citep{Soja2014, Soja2015}
and one VLBI session from 2017 \citep{Soja2018}.
Results of these works will be discussed in more detail in further sections
of this paper.

\section{Electron density models}
\subsection{Power-law model}

Typically, when it comes to measuring the electron density in the solar corona
by means of radio sounding,
most works tend to adopt a simple symmetric power-law model of the electron density
\begin{equation}
    \label{eq:powerlaw}
    \Ne(r) = \frac{N_0}{r^\alpha},
\end{equation}
where $N_0 = \Ne(\Rsun)$ is an estimated parameter with the physical
meaning of electron number density at the surface of the Sun and
$\alpha$ is either estimated or set to be approximately equal to 2.

A simplistic physical perspective gives a rough estimate of what
the value of $\alpha$ could or should be.
Suppose that the solar corona has a stationary electron density distribution
$\Ne(r)$ and that there is a radial outflow of the electrons
with non-constant drift velocity $v(r)$. Then the continuity equation must hold
\citep{Parker1958, Meyer-Vernet2007}:
\begin{equation}
    \frac{\partial \Ne(r)}{\partial t} + \nabla \cdot \vecc j(r) = 0,
\end{equation}
where $\vecc j(r) = \hat{\vecc r}\,v(r) \Ne(r)$ is the particle flux.
The density distribution is said to be stationary, therefore,
$\partial \Ne/ \partial t = 0$ and
\begin{equation}
    \frac{\partial}{\partial r}\left( r^2 v(r) \Ne(r) \right) = 0,
\end{equation}
\vspace{-2ex}
\begin{equation}
    r^2 v(r) \Ne(r) = \mathrm{const}.
\end{equation}
Hence, if the electrons drift away with constant velocity,
the electron density decreases proportionally to $1/r^2$.
The density should fall faster than $1/r^2$ if the electrons are accelerating
and slower if the electrons are decelerating.
As some works suggest \citep{Pierrard1999, Cranmer2007}, electrons might
indeed be accelerating on heliospheric distances up to roughly $40\Rsun$,
and in that case fitting the power-law density model to observational data
should naturally lead to powers greater than~2.

In fact, values of $N_0$ and $\alpha$ in \cref{eq:powerlaw}
vary significantly between different works
\citep{Berman1977}, which is usually attributed to variances in solar
activity. Moreover, they vary even between different sides of the Sun during
a single experiment  with a single spacecraft \citep{Anderson1987}, which
raises concern about the validity of symmetric models.

\begin{figure}[h]
    \centering
    \includegraphics[height=8cm]{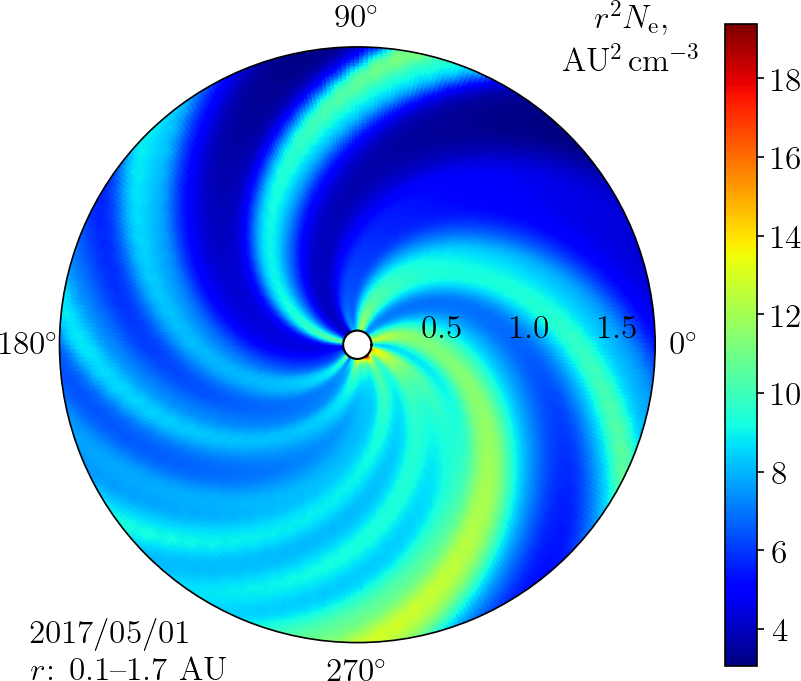}
    \caption{
        Solar wind electron density from the ENLIL model.
        The data is plotted in the plane with latitude of 0\degree\ in
        HEEQ coordinate system \citep[see][]{Thompson2006}.
    }
    \label{fig:enlil}
\end{figure}

\subsection{Three-dimensional MHD models}

A compelling alternative to symmetric power-law models is offered
by a class of numerical three-dimensional magnetohydrodynamic (MHD) models.
Unlike power-law models, which, at any given time,
are but an approximation
for real electron density distributions occurring in the solar corona,
3D MHD models aim to provide real-time spatial distributions
of electron density, velocity and other parameters of the solar plasma.
That is achieved by numerically solving a set of non-linear differential
equations called \emph{MHD equations} with boundary condition
for the magnetic field on solar surface given by synoptic magnetograms.

Since MHD models do not impose any restrictions on the solution,
which symmetric power-law models obviously do in multiple ways,
and also rely on physical principles and observational data,
they should, in principle, lead to substantially more accurate results
than the power law.
An illustration to this claim is given in \cref{fig:enlil},
which shows an electron density distribution given by a 3D MHD model
and suggests that the solar corona is highly non-stationary and
non-symmetric and that symmetric models of any kind,
let alone the power law, are not
sophisticated enough to accurately describe it.

Examples of MHD models are the ENLIL solar wind
model~\citep{Taktakishvili2011}, which provides solutions
for heliospheric distances greater than 0.1~AU ($21.5\Rsun$), and
Alfv\'en Wave Solar Model
\citep[AWSoM-R,\footnote{``R'' stands for ``real time''}][]{Sokolov2013,vanderholst2014},
implemented on top of the Space Weather Modeling Framework
\citep[SWMF, ][]{Toth2012}, which simulates
both inner and outer heliosphere on heliospheric distances
from $1.15\Rsun$ to $250\Rsun$.
Also available are results of validation of the AWSoM model
against observational data.
For instance,
\citet{vanderholst2014} have compared simulated
multi-wavelength extreme ultraviolet (EUV) images with observations from
Solar Terrestrial Relations Observatory (STEREO) and
Solar Dynamics Observatory (SDO);
\citet{Oran2013} have compared AWSoM model results with remote observations
in the extreme-ultraviolet and X-ray ranges from STEREO,
Solar and Heliospheric Observatory, and Hinode spacecraft and with
in situ measurements by Ulysses;
\citet{Moschou2018} have compared predictions
of the bremsstrahlung radio emission from the AWSoM model with observations
of the Sun at MHz and GHz frequencies.
All the studies have found the AWSoM model predictions to be in agreement
with observational data.

Simulation data for ENLIL and AWSoM models is provided by request on the CCMC
website.\footnote{\url{https://ccmc.gsfc.nasa.gov/}}

\vspace{1ex}
\subsection{Overview}

Generally speaking, the power-law model (i.e. $1/r^2$) is merely a typical
average radial profile of the electron density.
While it could, in theory, agree with some kind of observations
on a long-term scale (e.g. spacecraft ranging observations during
multiple months), it would be quite unreasonable to assume that it
would work well with instantaneous measures like VLBI.
Electron density at any fixed point in time and space cannot possibly
be accurately predicted by an average dependency.
Therefore, instantaneous observations should be in significantly
better agreement with non-symmetric time-dependent models like
ENLIL and AWSoM than with the power law.

\vspace{0.5ex}
\section{Radio wave propagation in the corona}

\subsection{Refractive index}

The dispersion relation for plasma is \citep{Goldston1995}
\begin{equation}
    \omega^2 = \omegap^2 + k^2 c^2,
\end{equation}
where $\omega$ and $k$ are the wave's angular frequency and wavevector,
$c$ is the speed of light, and $\omegap$ is the plasma frequency given by
\begin{equation}
    \omegap = \sqrt{\frac{\Ne \, e^2}{m_\mathrm{e} \epsilon_0}},
\end{equation}
where $\Ne(\vecc r)$ is the electron number density, $e$ and $m_\mathrm{e}$
are the electron's charge and mass, and $\epsilon_0$ is the electric constant.
The group velocity and the group refractive index are then
\begin{align}
    \vgroup &= \frac{\partial \omega}{\partial k}
            = c \, \sqrt{1 - \frac{\omegap^2}{\omega^2}}, \\
    \ngroup &= \frac{c}{\vgroup} = \frac{1}{\sqrt{1-\omegap^2/\omega^2}}.
\end{align}
Given that the frequency of VLBI X-band is about $10^9$~Hz and that
for heliospheric distances of $4\Rsun$, which is closer to the Sun
than observations by VLBI have ever been performed, the electron density
is about $10^{10}\ \mathrm{m}^{-3}$, the term $\omegap^2/\omega^2$
is not greater than $10^{-7}$.
Therefore, as $\omegap^2/\omega^2 \ll 1$, we can expand $\ngroup$
to first order of $\omegap^2/\omega^2$ and get
\begin{equation}
    \ngroup \approx 1 + \frac{\omegap^2}{2 \, \omega^2}.
\end{equation}
Rewriting the above in terms of critical plasma density
\begin{equation}
    \Ncrit(\omega) = \frac{m_\mathrm{e} \, \epsilon_0 \, \omega^2}{e^2}
        \approx 1.24 \cdot 10^{-2}\ \frac{\mathrm{s}^2}{\mathrm{m}^{3}} \cdot f^2
\end{equation}
and introducing spatial variability in the electron density, finally,
\begin{equation}
    \label{eq:n_group}
    \ngroup(\vecc r) \approx 1 + \frac{\Ne(\vecc r)}{2 \, \Ncrit(\omega)}
        \approx 1 + \frac{40.3\ \mathrm{m}^3\,\mathrm{s}^{-2}}{f^2} \Ne(\vecc r),
\end{equation}
where $f$ is the wave's frequency in Hz.

\subsection{Fermat's principle}

According to the Fermat's principle, the path taken between two points
by a photon is the path with the least optical path length (OPL)
\begin{equation}
    \label{eq:fermat}
    S = \int_{\vecc{r}_a}^{\vecc{r}_b} n \, \dd s
    \ \rightarrow \ \mathrm{min},
\end{equation}
where $n$ is the index of refraction, and the integral is taken along
the ray trajectory.

The above means that, in order to calculate the OPL between two points
in the solar corona, one has to solve the variational problem \eqref{eq:fermat},
taking into account the path curvature introduced by the Sun's gravity and plasma
\citep[see][chap.~8.3]{Perlick2000}.
However, calculations suggest that the variation of the OPL due to path curvature
is negligible and, therefore, taking the integral in \cref{eq:fermat}
along the straight line connecting points $\vecc r_a$ and $\vecc r_b$ gives
a reasonable approximation to the OPL.

\subsection{Dispersive time delay: symmetric power law model}

Thus, dispersive time delay in the solar corona can be written as
\begin{equation}
    \label{eq:tau_integral}
    \tau_\mathrm{cor} = \int_{\vecc r_a}^{\vecc r_b} (n_\mathrm{gr}(s, f) - 1) \, \dd s
               \approx \frac{1}{2 N_\mathrm{crit}(f)}\left[
                   \int_p^{r_a} \frac{\Ne(r)\,r \dd r}{\sqrt{r^2-p^2}} +
                   \int_p^{r_b} \frac{\Ne(r)\,r \dd r}{\sqrt{r^2-p^2}}
               \right],
\end{equation}
where $r_a = |\vecc r_a|$ and $r_b = |\vecc r_b|$,
and $p$ is the impact parameter, i.e. the distance of closest approach
of the light ray to the Sun.
The signs before the integrals in the above expression depend on the
spatial arrangement of $\vecc r_a$, $\vecc r_b$, and the Sun, with two “+” signs
corresponding to the case of $\vecc r_a$ and $\vecc r_b$ being on the opposite
sides of the Sun, which is always true for VLBI observations with small
elongation angles.
Substituting model \eqref{eq:powerlaw} into the above integral yields
an analytical expression:
\begin{equation}
    \label{eq:2f1_integral}
    \int_p^{x} \frac{r^{1-\alpha} \,\dd r}{\sqrt{r^2 - p^2}} =
    \frac{p^{1-\alpha}\sqrt{\pi}\,\Gamma{\left(\frac{\alpha-1}{2}\right)}}
         {2\,\Gamma{(\alpha/2)}}
    - \frac{x^{1-\alpha}}{\alpha - 1}
    \ _2F_1\left(\frac{1}{2},\frac{\alpha -1}{2};
                             \frac{\alpha+1}{2};
                             \frac{p^2}{x^2}\right),
\end{equation}
where $\Gamma(z)$ is the gamma function and
$_2F_1(a,b;c;z)$ is the hypergeometric function.

In the case of VLBI, $\vecc r_a$ and $\vecc r_b$ would be the radius vectors
of the radio telescope and the radio source.
Clearly, the heliospheric distance of the radio source is infinitely large,
which sets $x \rightarrow \infty$ in \cref{eq:2f1_integral}.
The improper form of the integral converges for $\alpha > 1$
as $\lim_{x \rightarrow \infty} x^{1-\alpha} = 0$.

\subsection{Dispersive time delay: numerical model}

The AWSoM model electron density data was obtained from the CCMC website, after
full model runs were ordered and performed for the dates of VLBI sessions.  For
each run, the website allows to download the electron density map on a uniform
grid in 3D spherical coordinates. From that map, the electron density
was interpolated onto a set of points on the straight line,
and the integral $\int_{\vecc{r}_a}^{\vecc{r}_b} (n-1) \, \dd s$
was computed numerically using Simpson's rule.
As the domain of the density map is finite, calculation
of the improper integral requires truncation at the domain border.

\section{Data}

\begin{table}[t]
    \centering
    \caption{List of VLBI sessions}
    \label{tbl:sessions}
    \begin{tabular}{lrrrr}
        \toprule
        \multicolumn{1}{c}{\bfseries Session} &
        \multicolumn{1}{c}{\bfseries Date} &
        \multicolumn{1}{c}{\bfseries No. of obs.}  &
        \multicolumn{1}{c}{\bfseries Minimum} &
        \multicolumn{1}{c}{\bfseries No. of obs.} \\
        & &  \multicolumn{1}{c}{\bfseries (< 15\degree/Total)} &
        \multicolumn{1}{c}{\bfseries elongation} &
        \multicolumn{1}{c}{\bfseries with $\tilde\tau_\mathrm{cor}$ > 1 cm\,$^\dagger$} \\
        \midrule
        RD1106 & 29 Nov. 2011 &    33/3695 &  4.0\degree &   11 \\
        RD1107 &  6 Dec. 2011 &    59/4242 &  4.1\degree &    8 \\
        RD1201 & 24 Jan. 2012 &    31/3482 &  4.9\degree &    9 \\
        RD1202 &  3 Apr. 2012 &    39/2776 &  5.8\degree &    7 \\
        RD1203 &  30 May 2012 &    52/2099 & 10.3\degree &    1 \\
        RD1204 & 19 June 2012 &     32/828 &  4.4\degree &    7 \\
        RD1205 & 10 July 2012 &   187/2953 &  6.0\degree &   34 \\
        RD1206 & 28 Aug. 2012 &   193/1558 &  3.8\degree &   60 \\
        RD1207 & 25 Sep. 2012 &   120/1727 &  6.1\degree &    5 \\
        RD1208 &  2 Oct. 2012 &   103/1918 &  3.9\degree &   15 \\
        RD1209 & 27 Nov. 2012 &    57/2731 &  4.3\degree &   22 \\
        RD1210 & 11 Dec. 2012 &    80/3540 &  4.8\degree &   12 \\
        AUA020 &   1 May 2017 &  1029/4010 &  1.2\degree &  945 \\
        AOV022 &   1 May 2018 & 3429/14099 &  1.3\degree & 3261 \\
        \bottomrule
        \multicolumn{5}{p{13cm}}{
            \footnotesize
            $^\dagger$\, $\tilde\tau_\mathrm{cor}$ is the differential delay
            calculated using model \eqref{eq:powerlaw} with
            $N_0 = 10^{12}$~m$^{-3}$, $\alpha = 2$.
            Number of observations with $\tilde\tau_\mathrm{cor}$ > 1~cm
            within a session can then be used as a measure of
            the session's sensitivity to the coronal electron density.
        }
    \end{tabular}
\end{table}

All VLBI sessions that have been previously used in other works
for solar corona parameters estimation
are listed in \cref{tbl:sessions}.
Those include 12 research and development (R\&D) sessions and
sessions AUA020 and AOV022.
The R\&D sessions were scheduled
by the IVS in 2011 and took place between November 2011 and December 2012.
The purpose of the R\&D sessions was to encourage the use of VLBI for
research related to general relativity and solar physics, therefore,
those sessions included observations of radio sources with elongations
as small as 4\degree, which is much less than the 15\degree\
elongation cutoff that was imposed in 2002 and kept until 2014.

\begin{figure}[t]
    \centering
    \includegraphics[height=7.5cm]{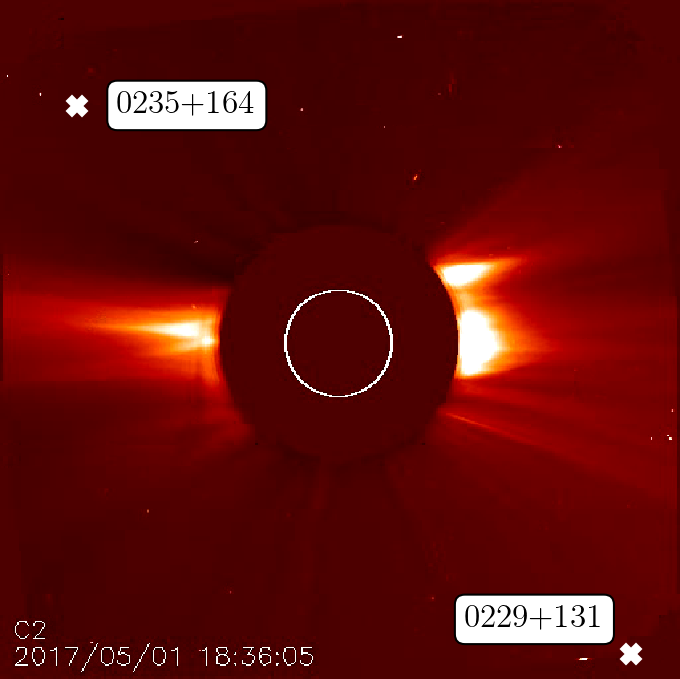}
    \hfill
    \includegraphics[height=7.5cm]{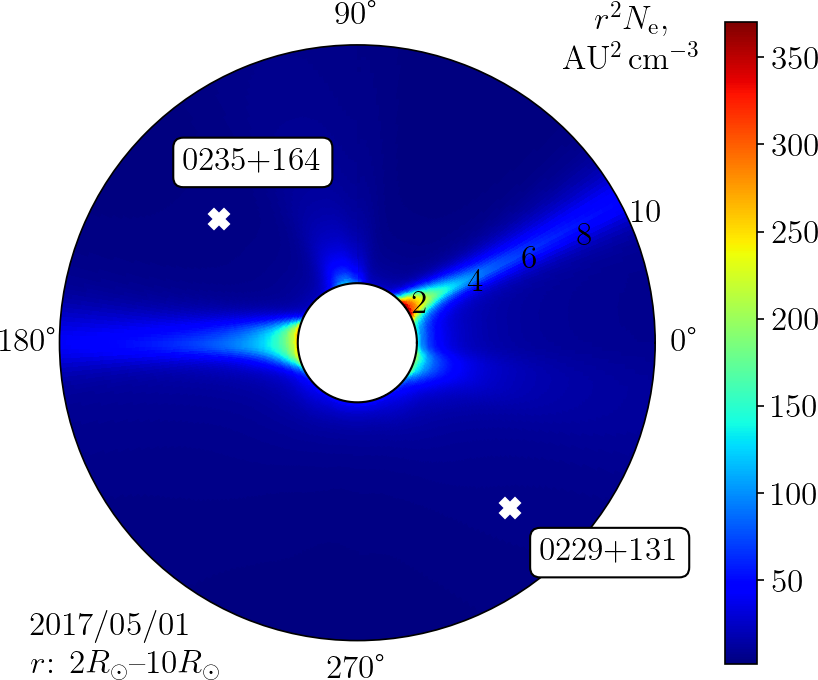}

    \caption{
        Observational geometry of the AUA020 session (as visible from Earth).
        LASCO C2 coronagraph image (left) and the solar corona electon density
        from the AWSoM model (right).
    }
    \label{fig:aua020_geometry}
\end{figure}

\begin{figure}[t]
    \centering
    \includegraphics[height=7.5cm]{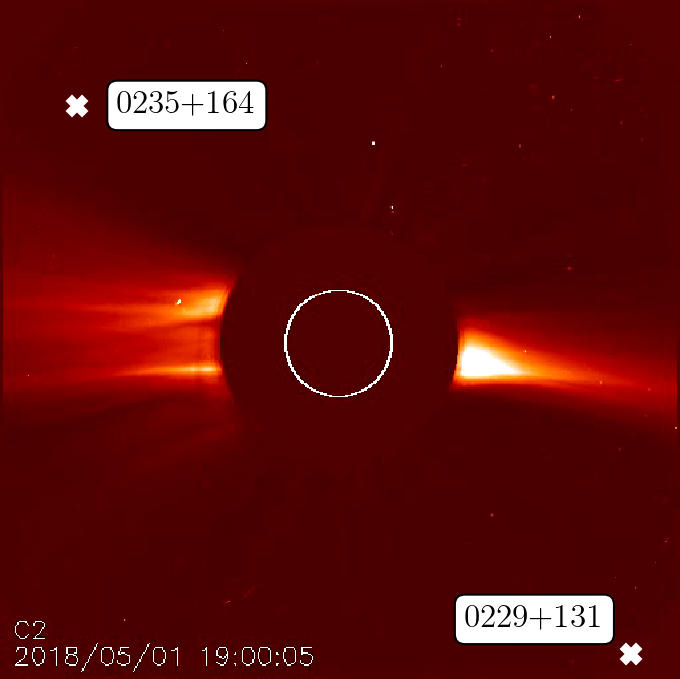}
    \hfill
    \includegraphics[height=7.5cm]{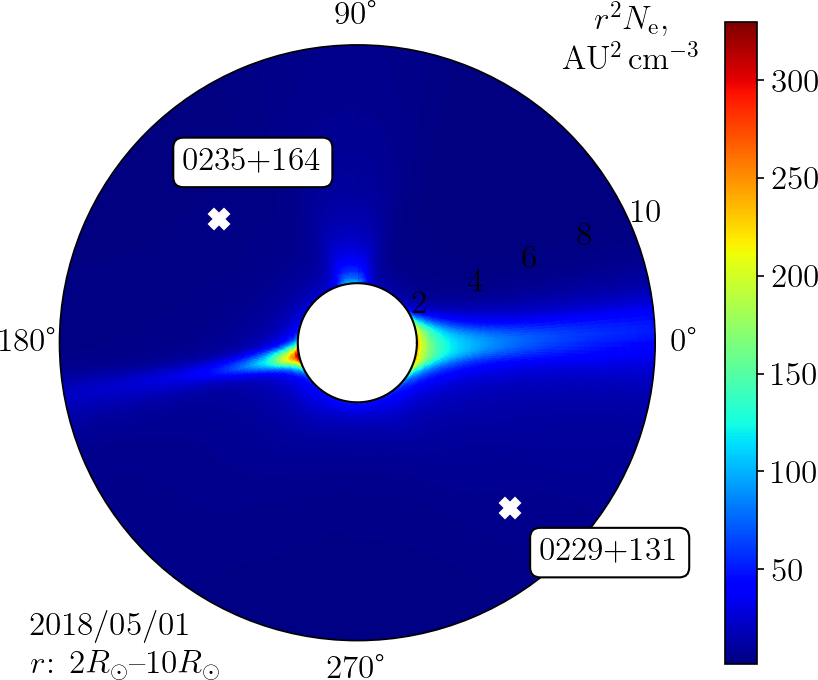}

    \caption{
        Observational geometry of the AOV022 session (as visible from Earth).
        LASCO C2 coronagraph image (left) and the solar corona electon density
        from the AWSoM model (right).
    }
    \label{fig:aov022_geometry}
\end{figure}

The AUA020 experiment was held on May 1, 2017 as part of the AUSTRAL
program with the purpose of testing general relativity \citep{Titov2018}.
Two radio sources that were close to the Sun were observed:
0235+164 (1.2\degree--1.5\degree) with 452 observations and
0229+131 (2.2\degree--2.7\degree) with 577 observations.
The AOV022 session (May 1, 2018) featured the same two radio sources:
0235+164 (1.3\degree--1.6\degree) with 1013 observations and
0229+131 (2.0\degree--2.3\degree) with 2416 observations.
Positions of the sources, coronagraph images of the solar
corona\,\footnote{LASCO C2 images were obtained online from
\url{https://cdaw.gsfc.nasa.gov/CME_list/}},
and AWSoM electron density maps for both sessions are shown
in \cref{fig:aua020_geometry,fig:aov022_geometry}.

The correlated VLBI data is available in the International VLBI Service
for Geodesy and Astrometry (IVS) data
archive\,\footnote{\url{https://ivscc.gsfc.nasa.gov/sessions/}}.

\needspace{6em}
\section{VLBI data analysis}

\subsection{General thoughts}

The quantity measured by VLBI is the group time delay between the signals
arriving at two telescopes. As the signals always pass through some dispersive
media with frequency-dependent refractive indexes, the total measured time delay
has a dispersive term $\tau_{\mathrm{disp}, f}$, which depends on the signal
frequency $f$.
Thus, VLBI measurements are performed at two frequencies:
S band ($\approx 2.3$~GHz) and X band ($\approx 8.4$~GHz) \citep{Sovers1991}.
If the delays measured at S and X band frequencies $f_\mathrm{s}$ and $f_\mathrm{x}$
are $\tau_\mathrm{s}$ and $\tau_\mathrm{x}$, then the expression for
the dispersive contribution to $\tau_\mathrm{x}$ can be easily derived
from \cref{eq:n_group}:
\begin{equation}
    \tau_\mathrm{disp, x} = \frac{f_\mathrm{s}^2}{f_\mathrm{s}^2-f_\mathrm{x}^2}
                            (\tau_\mathrm{x} - \tau_\mathrm{s}).
\end{equation}
Contribution of higher-order terms in the above expression would not be
greater than 0.5~mm \citep{Hawarey2005}, and, therefore, the first-order
approximation is said to be accurate enough.

The dispersive term of the time delay consists of delays in intergalactic
and interplanetary media, the Earth's ionosphere, the solar corona, and
receiver hardware. Delays in intergalactic and interplanetary media
can be considered negligible \citep[see][sec.~5]{Sovers1991}, and therefore
\begin{equation}
    \tau_\mathrm{disp, x} = \tau_\mathrm{cor} + \tau_\mathrm{ion}
                          + \tau_\mathrm{inst},
\end{equation}
where the three terms are the delays in the corona, the ionosphere, and
receiver hardware, respectively.

Taking $\Ne(r) = 0.5 \cdot 10^{12} / r^2$~m$^{-3}$
and VLBI baseline equal
to 12\,000~km and perpendicular to Earth---source direction,
we estimate that the coronal time delay should not be greater than 3~mm
for $p = 60\Rsun$, which corresponds to elongation of about 15\degree.
Therefore, elongation of 15\degree\ can be chosen as a threshold above which
the coronal time delay is undetectable by VLBI and is said to be equal to zero.

\subsection{Ionospheric delay}
\label{sec:ionosphere}

The ionospheric delay $\tau_\mathrm{ion}$ can be obtained either
by a least-squares procedure together with $\tau_\mathrm{cor}$ and
$\tau_\mathrm{inst}$ \citep[see][]{Soja2014} or independently by using
global ionosphere maps (GIMs), as done in \citep{Soja2015},
which are constructed from GPS satellite data. In this work, the
second approach was followed.
GIMs are routinely produced by the Center for Orbit Determination in Europe (CODE)
Analysis Center \citep{Schaer1996,CODE} and available via anonymous FTP server
of the Astronomical Institute of
University of Bern\,\footnote{\url{ftp://ftp.aiub.unibe.ch/CODE}}.
Ionospheric model is represented by a single infinitely thin layer corresponding
to the ionosphere layer F2 which has the largest charged particles density. Daily GIM
files contain vertical total electron content ($E_{v}$) in TEC units (TECU);
one TECU corresponds to $10^{16}$~m$^{-2}$ free electrons.

In the dual frequency VLBI observations the center frequency of X band is large compared
with the local plasma frequency, therefore the contribution of ionosphere to the group
delay \citep{Ros2000} can be determined as
\begin{equation}
    \tau_\mathrm{ion} = \frac{ 40.3\ \mathrm{m}^3\,\mathrm{s}^{-2} }{ c \, f^2 } \cdot E,
\end{equation}
where $E$ is the total electron content along the line of sight from the radio telescope
towards the target source. $E$ can be calculated from the GIM files taking into account
the height of the model layer $H = 450$~km above the mean Earth's surface radius
$R_0 = 6371$~km and applying the cosine mapping function
\begin{equation}
    E = \frac{E_{v}}{\cos \arcsin\left(\frac{R}{R_0 + H} \cos h\right) },
\end{equation}
where $R$ is radius with respect to the radio telescope position and
$h$ is the elevation angle of antenna pointing to the target source.

\needspace{6em}
\subsection{Data selection}

\begin{figure}[t]
    \centering
    \includegraphics{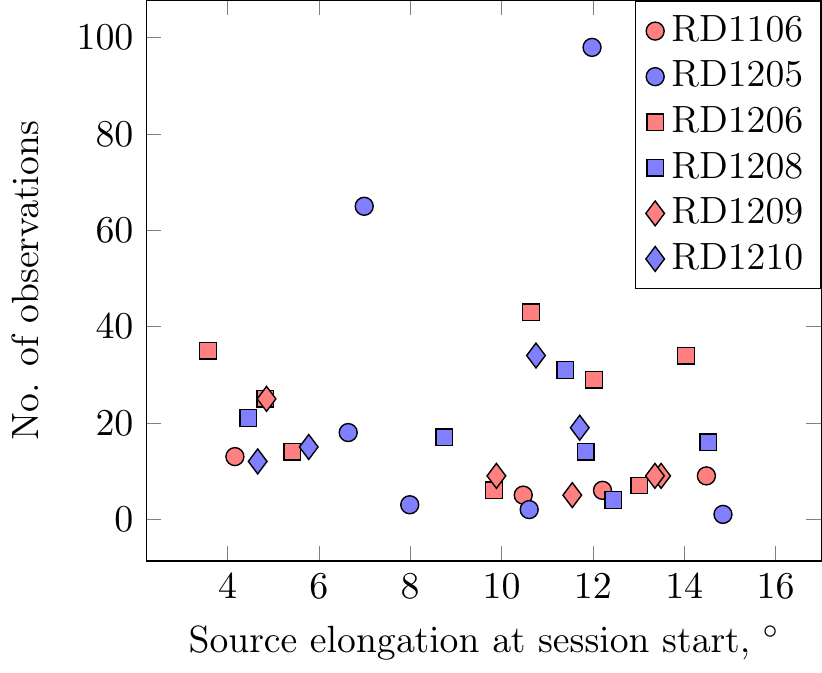}
    \caption{
        Elongation distribution in R\&D sessions
        (observations at elongations above $15\degree$ are excluded).
    }
    \label{fig:angles}
\end{figure}

Out of 12 R\&D sessions we picked the most sensitive to the solar corona
by using a criterion requiring a session to have more
than 10 observations with $\tilde\tau_\mathrm{cor} > 1$~cm,
where $\tilde\tau_\mathrm{cor}$ is calculated using power-law model
\eqref{eq:powerlaw} with $N_0 = 10^{12}$~m$^{-3}$, $\alpha = 2$.
Thus, six out of 12 sessions were taken, namely,
RD1106, RD1205, RD1206, RD1208, RD1209, and RD1210.
The reason for such filtering is that knowing the minimum elongation
for a session or even the numbers of observations for all elongations
is not enough to assert sensitivity of the data to $N_0$.
As the coronal delay depends not only on the source's elongation, but
also on the positions of the telescopes, some sort of filtering with respect
to the model delay values has to be performed.
Elongation distributions for the 6 picked R\&D sessions are shown
in \cref{fig:angles}.

\subsection{Algorithm}

We start by taking the correlated data and subtracting the ionospoheric delays
from the total delays using the procedure described in sec.~\ref{sec:ionosphere}.
Coordinates of stations and sources are fixed to the values
provided in NGS data since possible inaccuracies in the coordinates
do not affect the differential delay at any significant level.
To avoid correlations between the determined solar corona parameter
and instrumental biases,
we divide observations in each session into two groups:
the first group with elongation greater than $15\degree$ and the second
group with elongation less than $15\degree$.
The two groups are then used in two separate solutions.
In the first solution, instrumental per-station biases are determined,
which define $\tau_\mathrm{inst}$, and in the second solution, the
electron density multiplier $N_0$ is determined. Both adjustments are performed
with weighted linear least squares procedure, which, in the case
of the solution for instrumental biases, also includes iterative
outlier removal by the $3\sigma$ rule. In the solution for $N_0$
no outlier removal is performed.
Coronal delays are calculated by evaluating \cref{eq:tau_integral}
either using the analytical expression \eqref{eq:2f1_integral}
(for the power-law model), or numerically (for numerical electron density maps).

We performed least-squares estimations for both the power law and
the AWSoM model to see whether VLBI is able to assert the AWSoM model's
superiority over the power law.
It is already known from the studies carried out
by \citet{Soja2014,Soja2015,Soja2018}
that VLBI observations are sensitive
enough to allow for estimation of the parameters of the power-law electron
density model.
However, under the assumption of the power law being inaccurate and
the true electron density being described by MHD models, there is still
unclarity regarding the degree to which VLBI observations
can be utilized in measuring real non-symmetric electron density
in the corona and comparing different models.
In the following section we will show
that non-stationary and non-symmetric artifacts seen in MHD solutions
can indeed make a noticeable impact on VLBI observations' residuals.

\section{Results}
\label{results}

\begin{table}[t]
    \centering
    \caption{Results}
    \label{tbl:results}
    \begin{tabular}{lrrrrrrr}
        \toprule
        \multicolumn{1}{c}{\bfseries Session} &
        \multicolumn{1}{c}{\bfseries Outliers} &
        \multicolumn{1}{c}{\bfseries $\boldsymbol\alpha$} &
        \multicolumn{1}{c}{\bfseries Soja et al.$^\dagger$} &
        \multicolumn{2}{c}{\bfseries Power law$^{\dagger\dagger}$} &
        \multicolumn{2}{c}{\bfseries AWSoM}\\
        & & &
        \multicolumn{1}{c}{ \bfseries $\boldsymbol{N_0}$ ($\mathbf{10^{12}}$ m$^{-3})$} &
        \multicolumn{1}{c}{\bfseries $\boldsymbol{N_0}$ ($\mathbf{10^{12}}$ m$^{-3})$} &
        \multicolumn{1}{c}{\bfseries RMS (m)} &
        \multicolumn{1}{c}{\bfseries $A$} &
        \multicolumn{1}{c}{\bfseries RMS (m)} \\
        \midrule
        RD1106 &  114 &   2 &   $0.0\pm0.4$  & $0.10\pm0.47$ & 0.0485 & $-1.43\pm1.28$ & 0.0476 \\
        RD1205 &  214 &   2 &   $0.5\pm0.3$  & $1.08\pm0.21$ & 0.0211 & $ 0.21\pm0.12$ & 0.0223 \\
        RD1206 &   98 &   2 &   $0.3\pm0.1$  & $0.34\pm0.15$ & 0.0288 & \ellps & \ellps \\
        RD1208 &   56 &   2 &   $1.5\pm0.4$  & $0.86\pm0.20$ & 0.0315 & \ellps & \ellps \\
        RD1209 &  234 &   2 &   $0.1\pm0.3$  & $0.48\pm0.25$ & 0.0497 & \ellps & \ellps \\
        RD1210 &   51 &   2 &   $2.5\pm0.6$  & $0.00\pm0.68$ & 0.0482 & $ 0.93\pm0.80$ & 0.0478 \\
        AUA020 &  155 & 2.2 & $0.61\pm0.05$  & $0.57\pm0.01$ & 0.0655 & $ 0.96\pm0.02$ & 0.0612 \\
        \multirow{2}{*}{AOV022} & \multirow{2}{*}{1203} &   2 & $0.44\pm0.005$ & $0.43\pm0.02$ & 0.15607 &
                \multirow{2}{*}{$0.84\pm0.04$} & \multirow{2}{*}{0.1530} \\
        & &   2.3 & \ellps & $0.60\pm0.03$ & 0.15604 && \\
        \bottomrule
        \multicolumn{8}{p{17cm}}{
            \footnotesize
            $^{\phantom{\dagger}\dagger}$
            Values were taken from \citep{Soja2014} for R\&D sessions,
            from \citep{Soja2018} for AUA020, and from \citep{Soja2019} for AOV022.

            $^{\dagger\dagger}$ For the R\&D sessions, the default value $\alpha=2$
            was used. As discussed in sec.~\ref{results},
            residuals for these sessions do not have clear dependencies
            on the value of $\alpha$, and therefore it is not possible
            to justify a different choice of $\alpha$.
            For sessions AUA020 and AOV022, data shown in \cref{fig:rmss}
            states that residuals do reach minimums at 2.2 and 2.3,
            respecively, so those powers were used.
            Power 2 for AOV022 is given only for the purpose of comparison
            with \citep{Soja2019}, which gives values of $N_0$ only for
            powers 2 and 2.74.
        }
    \end{tabular}
\end{table}

\begin{figure}[t]
    \centering
    \includegraphics{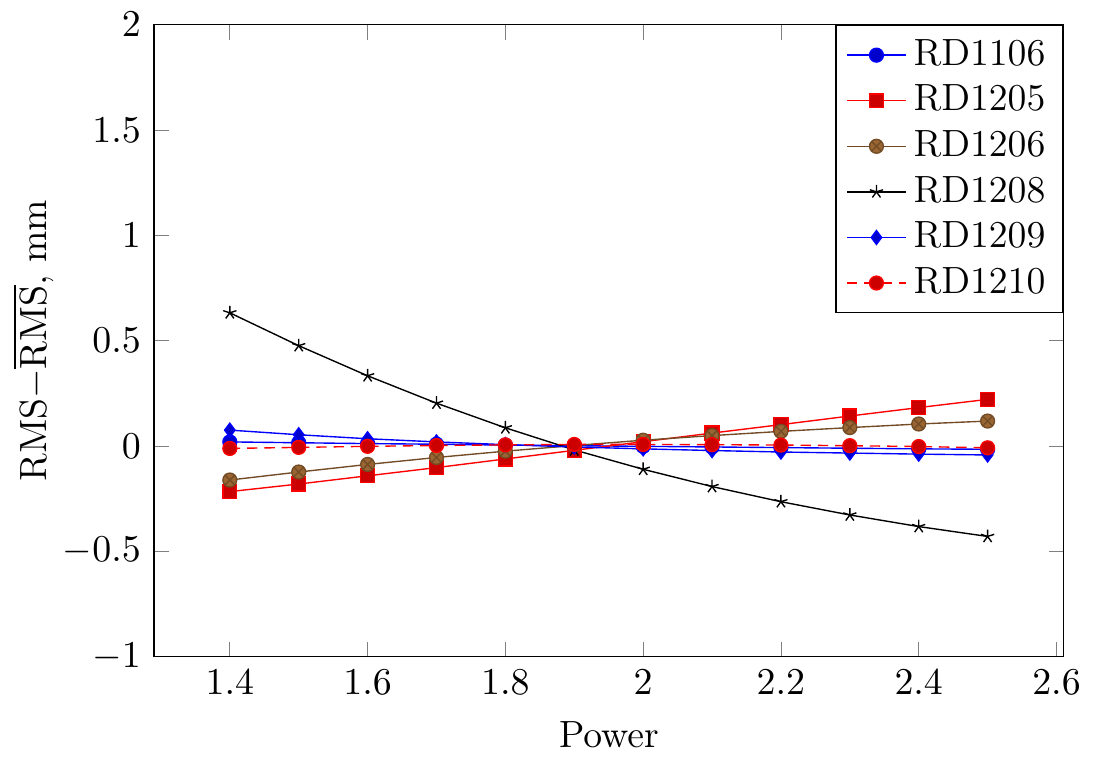}
    \includegraphics{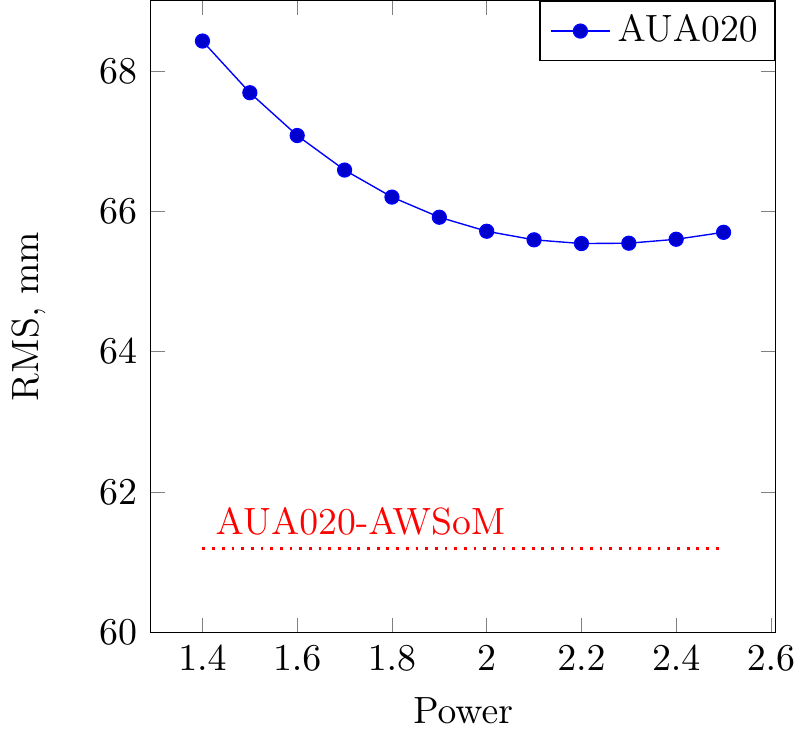}
    \includegraphics{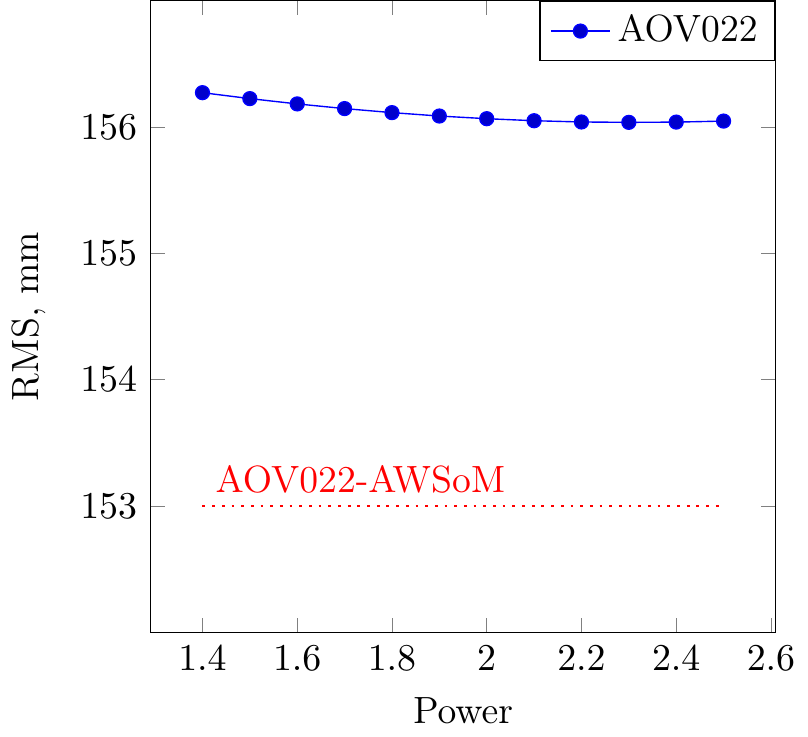}
    \caption{
        Dependencies of root-mean-squared (RMS) errors on the power $\alpha$
        of the power-law electron density model.
        For each value of $\alpha$ the value of $N_0$ is optimal, i.e.
        acquired via least squares.
        In the left picture, each RMS curve's average value was subtracted from
        the curve to clearly show dynamics of the residuals.
        The red dotted lines in the bottom pictures correspond to RMS errors
        of 0.0622~m and 0.154~m resulting from processing sessions
        AUA020 and AOV022 with electron density maps produced by
        the AWSoM model.
        }
    \label{fig:rmss}
\end{figure}

Solutions with the symmetric power-law model \eqref{eq:powerlaw} were acquired
for all sessions.
During sessions RD1206, RD1208, and RD1209, strong coronal mass ejections (CMEs)
happened, which made it impossible to apply the AWSoM data to these sessions.
Because of that, solutions with the AWSoM model were obtained for all
sessions except RD1206, RD1208, and RD1209.
In the case of the AWSoM model,
the estimated parameter was the unitless multiplier $A$ of the electron
density map. The model was run with default parameters, which, as the authors
have noted, need further calibration and are not guaranteed to be the best.
Thus, the multiplier $A$ was artificially introduced to compensate,
to some extent, for the uncertainties of the model input parameters.
Values of $N_0$ and $A$ and their uncertainties ($1\sigma$) for each session
are given in \cref{tbl:results}.

The values of $N_0$ in \cref{tbl:results} generally agree within
the uncertainty with the ones provided by \citet{Soja2014,Soja2015,Soja2018},
and agreement is almost perfect for AUA020 and, in case of $\alpha = 2$, AOV022.
The only session with considerable difference in solutions is RD1210,
where Soja's result is $2.5\pm0.6$ and ours is $0.00\pm0.68$.
In fact, taking the uncertainty into account, even these results
agree within $3\sigma$.
Note that the zero value of $N_0$ should be treated not as an actual zero
electron density but as an artifact of the least-squares solution.
Also, the uncertainty is large enough to cover a broad range
of physically reasonable positive values of $N_0$.

Although least-squares estimations have succeeded at finding
solutions for $N_0$ and confirming the results obtained independently
by \citet{Soja2014,Soja2018,Soja2019},
still, there is no evidence that the R\&D sessions
are sensitive to the choice of the solar corona model and
to the parameters of the chosen model.
Not only the $N_0$ uncertainties for the R\&D sessions are somewhat large,
but, as shown in \cref{fig:rmss}, the post-fit residuals barely depend
on the choice of the power $\alpha$, varying by less than 0.5--1 mm
with $\alpha$ changing from 1.4 to 2.5 and not even reaching minimum values
for any given $\alpha$.
Hence, from the perspective of the R\&D sessions data,
all possible power laws fit the electron density equally well, including those
with extreme $\alpha$ values that never occur in previously published papers
and therefore seem unlikely in general. Furthermore, the residuals for
the R\&D sessions with the AWSoM model (given in \cref{tbl:results})
differ only slightly from those for the power law
and fail to show any consistent relationship to them.

To put it differently, we are faced with an ultimate impossibility
to reliably determine \emph{any} solar corona parameters from
the R\&D sessions. The power $\alpha$ is indeterminable
since for virtually any $\alpha$ there exists an optimal $N_0(\alpha)$ which
fits the observations almost equally well as any other optimal
$(\alpha,\,N_0(\alpha))$ pair.
And even if we assume $\alpha$ to be fixed and accurate,
the estimated values of $N_0$ differ greatly between different sessions
and uncertainties cover wide ranges of $N_0$,
making the solutions reliable only in a purely mathematical sense
and disconnected from physical reality.

A session that appears to be sensitive to the electron density models is AUA020.
The reasons for that have to be (i) a fairly large number of observations
close to the Sun and (ii) outstandingly small elongation angles.
The uncertainty of $N_0$ is by an order of magnitude
smaller than the uncertainties for the R\&D sessions, and, importantly,
the dependency of residuals on the power $\alpha$ (\cref{fig:rmss})
clearly shows a minimum at $\alpha \approx 2.2$.
With the AWSoM model, the residuals for AUA020 decrease by $\approx 4$~mm,
which means that the model eliminates a scatter of 2~cm in root-mean-square sense
and therefore might be considered to be a significant improvement over
the power-law model.

The situation is quite similar with the AOV022 session, which also shows
a residuals minimum at $\alpha \approx 2.2$, sufficiently small
estimated parameter uncertainties, and a 3~mm decrease in residuals with
the AWSoM model.

\section{Conclusion and future work}

We have shown that the VLBI technique is capable of providing observational
data sensitive enough to the solar corona electron density to distinguish between
different models and to allow for estimation of model parameters.
However, VLBI sessions need to be planned carefully to include a large number
of observations at small heliocentric distances,
since VLBI's sensitivity to the solar corona decreases severely
at elongations even as large as 4\degree, as we have seen in the case
of R\&D sessions.

Having confirmed recent results of \citep{Soja2014, Soja2015, Soja2018},
we also note that the AUA020 and AOV022 sessions show superiority of the AWSoM model over
the power law, and suggest that the power law is not an accurate
representation of the solar corona electron density and that further analysis
can significantly benefit from more sophisticated modern 3D solar corona
models. The AWSoM model results were successfully compared to different kinds
of observations in multiple works \citep[see e.g.][]{Sokolov2013,Oran2013,Moschou2018};
VLBI observations of quasars provide an additional source of validation.
A more careful analysis is needed in presence of strong coronal mass
ejections (CMEs), where AWSoM data suggests extremely high electron density
in certain regions. Currently, the available VLBI observations
do not fit to the numerical model with strong CMEs as well as they
do with a symmetric model.

The R\&D sessions have proved to be unable to determine anything
but the mere fact that the solar corona exists and that its electron density decreases
with heliospheric distance, which is already a universally accepted truth.
To fully explore the possibilities that VLBI provides for
comparing different solar corona models, there has to be more
observational data available, preferably of the likes of AUA020 and AOV022
in terms of the number of observations made close to the Sun
and minimum elongation angles. As~\cite{Titov2018} point out,
that requires strong radio sources being close to the Sun,
and also large radio telescopes and high data recording rate.

\section*{Acknowledgements}

Authors would like to thank their colleague Pavel Volkov for clarification of
some concepts related to the solar plasma physics, and Steven R. Cranmer from the
University of Colorado, Boulder for useful advice.

Authors are grateful to the IVS~\citep{IVS} for coordinating the VLBI observations and
providing the correlated data, and to the CODE Analysis Center in the
Astronomical Institute of University of Bern for the global ionosphere
map data.

AUA020 and AOV022 sessions were performed by different VLBI stations across the
world, including those of the Australian VLBI network~\citep{AUSTRAL}, Russian
``QUASAR'' VLBI network~\citep{Quasar2019}, HartRAO in South
Africa~\citep{Mayer2014}, and also stations from Japan (Ishioka), China
(Sheshan, Kunming), and South Korea (Sejong).

Solar corona simulation results have been provided by
the Community Coordinated Modeling Center at Goddard Space Flight Center
through their public Runs on Request system
(\url{http://ccmc.gsfc.nasa.gov/}).
The following CCMC runs were requested (the data is available
at \url{http://ccmc.gsfc.nasa.gov/ungrouped/SH/Solar_main.php}):
\textrm{Dan\_\allowbreak Aksim\_\allowbreak 062019\_SH\_3} for RD1106,
\textrm{Dmitry\_\allowbreak Pavlov\_\allowbreak 080119\_SH\_1} for RD1205,
\textrm{Dmitry\_\allowbreak Pavlov\_\allowbreak 061919\_SH\_1} for RD1210,
\textrm{Dmitry\_\allowbreak Pavlov\_\allowbreak 052919\_SH\_1} for AUA020,
\textrm{Dan\_\allowbreak Aksim\_\allowbreak 062619\_SH\_1 } for AOV022.

The ENLIL model was developed by the J.~Linker, Z.~Mikic, R.~Lionello, P.~Riley,
N.~Arge, and D.~Odstrcil at the University of Colorado.
The Space Weather Modeling Framework (SWMF) was developed by the Center
for Space Environment Modeling (CSEM) team led by Tamas Gombosi at
the University of Michigan \citep{Toth2012}.

The SOHO/LASCO data used here are produced by a consortium of the Naval Research
Laboratory (USA), Max-Planck-Institut f\"{u}r Sonnensystemforschung (Germany),
Laboratoire d'Astrophysique Marseille (France), and the University of Birmingham
(UK). SOHO is a project of international cooperation between ESA and NASA.

\bibliography{references}

\begin{thebibliography}{}
\expandafter\ifx\csname natexlab\endcsname\relax\def\natexlab#1{#1}\fi
\providecommand{\url}[1]{\href{#1}{#1}}
\providecommand{\dodoi}[1]{doi:~\href{http://doi.org/#1}{\nolinkurl{#1}}}
\providecommand{\doeprint}[1]{\href{http://ascl.net/#1}{\nolinkurl{http://ascl.net/#1}}}
\providecommand{\doarXiv}[1]{\href{https://arxiv.org/abs/#1}{\nolinkurl{https://arxiv.org/abs/#1}}}

\bibitem[{Anderson {et~al.}(1987)Anderson, Krisher, Borutzki, Connally, Eshe,
  Hotz, Kinslow, Kursinski, Light, Matousek, Moyd, Roth, Sweetnam, Taylor,
  Tyler, Gresh, \& Rosen}]{Anderson1987}
Anderson, J.~D., Krisher, T.~P., Borutzki, S.~E., {et~al.} 1987, The
  Astrophysical Journal, 323, L141, \dodoi{10.1086/185074}

\bibitem[{Bale {et~al.}(2016)Bale, Goetz, Harvey, Turin, Bonnell,
  Dudok de Wit, Ergun, MacDowall, Pulupa, Andre, Bolton, Bougeret, Bowen,
  Burgess, Cattell, Chandran, Chaston, Chen, Choi, Connerney, Cranmer,
  Diaz-Aguado, Donakowski, Drake, Farrell, Fergeau, Fermin, Fischer, Fox,
  Glaser, Goldstein, Gordon, Hanson, Harris, Hayes, Hinze, Hollweg, Horbury,
  Howard, Hoxie, Jannet, Karlsson, Kasper, Kellogg, Kien, Krasnoselskikh,
  Krucker, Lynch, Maksimovic, Malaspina, Marker, Martin, Martinez-Oliveros,
  McCauley, McComas, McDonald, Meyer-Vernet, Moncuquet, Monson, Mozer, Murphy,
  Odom, Oliverson, Olson, Parker, Pankow, Phan, Quataert, Quinn, Ruplin, Salem,
  Seitz, Sheppard, Siy, Stevens, Summers, Szabo, Timofeeva, Vaivads, Velli,
  Yehle, Werthimer, \& Wygant}]{Bale2016}
Bale, S.~D., Goetz, K., Harvey, P.~R., {et~al.} 2016, Space Science Reviews,
  204, 49, \dodoi{10.1007/s11214-016-0244-5}

\bibitem[{{Berman}(1977)}]{Berman1977}
{Berman}, A.~L. 1977, {Electron Density in the Extended Corona: Two Views}, The
  deep space network progress report 42-41, NASA JPL

\bibitem[{{Bird} \& {Edenhofer}(1990)}]{Bird1990}
{Bird}, M.~K., \& {Edenhofer}, P. 1990, {Remote Sensing Observations of the
  Solar Corona}, ed. R.~{Schwenn} \& E.~{Marsch} (Springer Berlin Heidelberg),
  13--97

\bibitem[{Bird {et~al.}(2012)Bird, Pätzold, Häusler, Asmar, Tellmann, Hahn,
  Efimov, \& Chashei}]{Bird2012}
Bird, M.~K., Pätzold, M., Häusler, B., {et~al.} 2012, in 511th
  WE-Heraeus-Seminar, Physikzentrum Bad Honnef, Germany.
\newblock \url{http://www2.mps.mpg.de/meetings/hcor/p/p01.pdf}

\bibitem[{Cranmer {et~al.}(2007)Cranmer, A.~van Ballegooijen, \&
  Edgar}]{Cranmer2007}
Cranmer, S., A.~van Ballegooijen, A., \& Edgar, R. 2007, The Astrophysical
  Journal Supplement Series, 171, \dodoi{10.1086/518001}

\bibitem[{Cranmer(2002)}]{Cranmer2002}
Cranmer, S.~R. 2002, Space Science Reviews, 101, 229,
  \dodoi{10.1023/A:1020840004535}

\bibitem[{Dach {et~al.}(2018)Dach, Schaer, J\"{a}ggi, Prange, Stebler, Sidorov,
  Arnold, \& Villiger}]{CODE}
Dach, R., Schaer, S., J\"{a}ggi, A., {et~al.} 2018, CODE final product series
  for the IGS,  Astronomical Institute, University of Bern,
  \dodoi{10.7892/boris.75876.3}.
\newblock \url{https://boris.unibe.ch/119490/}

\bibitem[{Goldston \& Rutherford(1995)}]{Goldston1995}
Goldston, R.~J., \& Rutherford, P.~H. 1995, Introduction to Plasma Physics
  (Bristol, UK Philadelphia: Institute of Physics Publishing)

\bibitem[{Harrison {et~al.}(2008)Harrison, Davis, Eyles, Bewsher, Crothers,
  Davies, Howard, Moses, Socker, Newmark, Halain, Defise, Mazy, Rochus, Webb,
  \& Simnett}]{Harrison2008}
Harrison, R.~A., Davis, C.~J., Eyles, C.~J., {et~al.} 2008, Solar Physics, 247,
  171, \dodoi{10.1007/s11207-007-9083-6}

\bibitem[{Hawarey {et~al.}(2005)Hawarey, Hobiger, \& Schuh}]{Hawarey2005}
Hawarey, M., Hobiger, T., \& Schuh, H. 2005, Geophysical Research Letters, 32,
  \dodoi{10.1029/2005GL022729}

\bibitem[{Kelso(1959)}]{Kelso1959}
Kelso, J. 1959, Journal of Atmospheric and Terrestrial Physics, 16, 357,
  \dodoi{10.1016/0021-9169(59)90085-6}

\bibitem[{Mayer {et~al.}(2014)Mayer, B\"{o}hm, Combrinck, Botai, \&
  B\"{o}hm}]{Mayer2014}
Mayer, D., B\"{o}hm, J., Combrinck, L., Botai, J., \& B\"{o}hm, S. 2014, Acta
  Geodaetica et Geophysica, 49, 313, \dodoi{10.1007/s40328-014-0063-7}

\bibitem[{Meyer-Vernet(2007)}]{Meyer-Vernet2007}
Meyer-Vernet, N. 2007, Basics of the Solar Wind, Cambridge Atmospheric and
  Space Science Series (Cambridge University Press),
  \dodoi{10.1017/CBO9780511535765}

\bibitem[{Moschou {et~al.}(2018)Moschou, Sokolov, Cohen, Drake, Borovikov,
  Kasper, Alvarado-Gomez, \& Garraffo}]{Moschou2018}
Moschou, S.-P., Sokolov, I., Cohen, O., {et~al.} 2018, The Astrophysical
  Journal, 867, 51, \dodoi{10.3847/1538-4357/aae58c}

\bibitem[{Nothnagel {et~al.}(2017)Nothnagel, Artz, Behrend, \& Malkin}]{IVS}
Nothnagel, A., Artz, T., Behrend, D., \& Malkin, Z. 2017, Journal of Geodesy,
  91, 711, \dodoi{10.1007/s00190-016-0950-5}

\bibitem[{Oran {et~al.}(2013)Oran, van~der Holst, Landi, Jin, Sokolov, \&
  Gombosi}]{Oran2013}
Oran, R., van~der Holst, B., Landi, E., {et~al.} 2013, The Astrophysical
  Journal, 778, 176, \dodoi{10.1088/0004-637x/778/2/176}

\bibitem[{Parker(1958)}]{Parker1958}
Parker, E.~N. 1958, The Astrophysical Journal, 128, 664, \dodoi{10.1086/146579}

\bibitem[{Perlick(2000)}]{Perlick2000}
Perlick, V. 2000, Ray Optics, Fermat’s Principle, and Applications to General
  Relativity, Lecture Notes in Physics Monographs (Springer)

\bibitem[{Pierrard {et~al.}(1999)Pierrard, Maksimovic, \&
  Lemaire}]{Pierrard1999}
Pierrard, V., Maksimovic, M., \& Lemaire, J. 1999, Journal of Geophysical
  Research: Space Physics, 104, 17021, \dodoi{10.1029/1999JA900169}

\bibitem[{Plank {et~al.}(2017)Plank, Lovell, McCallum, Mayer, Reynolds, Quick,
  Weston, Titov, Shabala, B{\"o}hm, Natusch, Nickola, \& Gulyaev}]{AUSTRAL}
Plank, L., Lovell, J. E.~J., McCallum, J.~N., {et~al.} 2017, Journal of
  Geodesy, 91, 803, \dodoi{10.1007/s00190-016-0949-y}

\bibitem[{{Ros} {et~al.}(2000){Ros}, {Marcaide}, {Guirado}, {Sard{\'o}n}, \&
  {Shapiro}}]{Ros2000}
{Ros}, E., {Marcaide}, J.~M., {Guirado}, J.~C., {Sard{\'o}n}, E., \& {Shapiro},
  I.~I. 2000, Astronomy and Astrophysics, 356, 357

\bibitem[{Schaer {et~al.}(1996)Schaer, Beutler, Rothacher, \&
  Springer}]{Schaer1996}
Schaer, S., Beutler, G., Rothacher, M., \& Springer, T.~A. 1996, in Proceedings
  of the IGS Analysis Center Workshop 1996, ed. R.~Neilan, P.~Van~Scoy, \&
  J.~Zumberge, 181--192

\bibitem[{Schuh \& Behrend(2012)}]{Schuh2012}
Schuh, H., \& Behrend, D. 2012, Journal of Geodynamics, 61, 68,
  \dodoi{10.1016/j.jog.2012.07.007}

\bibitem[{Shuygina {et~al.}(2019)Shuygina, Ivanov, Ipatov, Gayazov, Marshalov,
  Melnikov, Kurdubov, Vasilyev, Ilin, Skurikhina, Surkis, Mardyshkin,
  Mikhailov, Salnikov, Vytnov, Rakhimov, Dyakov, \& Olifirov}]{Quasar2019}
Shuygina, N., Ivanov, D., Ipatov, A., {et~al.} 2019, Geodesy and Geodynamics,
  10, 150, \dodoi{10.1016/j.geog.2018.09.008}

\bibitem[{Soja {et~al.}(2014)Soja, Heinkelmann, \& Schuh}]{Soja2014}
Soja, B., Heinkelmann, R., \& Schuh, H. 2014, Nature Communications, 5,
  \dodoi{10.1038/ncomms5166}

\bibitem[{Soja {et~al.}(2015)Soja, Heinkelmann, \& Schuh}]{Soja2015}
Soja, B., Heinkelmann, R., \& Schuh, H. 2015, in IAG 150 Years (Cham: Springer
  International Publishing), 611--616

\bibitem[{Soja {et~al.}(2018)Soja, Titov, Girdiuk, Mccallum, Shabala, Mccallum,
  Mayer, Schartner, {Alet De Witt}, {Fengchun Shu}, Melnikov, Ivanov,
  Mikhailov, {Sang-Oh Yi}, Lambert, Kawai, \& Haas}]{Soja2018}
Soja, B., Titov, O., Girdiuk, A., {et~al.} 2018, Solar Corona Electron Density
  Models from Recent VLBI Experiments AUA020 and AUA029,
  \dodoi{10.13140/rg.2.2.21438.38728}

\bibitem[{Soja {et~al.}(2019)Soja, Heinkelmann, {H. Schuh}, Böhm, Sun, Titov,
  Lovell, McCallum, Shabala, McCallum, Mayer, Schartner, Witt, {Fengchun Shu},
  {B. Xia}, {T Jiang}, Melnikov, Ivanov, Mikhailov, {Sangoh Yi}, Kawai, Haas,
  \& {Yd He}}]{Soja2019}
Soja, B., Heinkelmann, R., {H. Schuh}, {et~al.} 2019, Very Long Baseline
  Interferometry as a tool to probe the solar corona,  Unpublished,
  \dodoi{10.13140/rg.2.2.17735.04009}.
\newblock \url{http://rgdoi.net/10.13140/RG.2.2.17735.04009}

\bibitem[{Sokolov {et~al.}(2013)Sokolov, van~der Holst, Oran, Downs, Roussev,
  Jin, Manchester, Evans, \& Gombosi}]{Sokolov2013}
Sokolov, I.~V., van~der Holst, B., Oran, R., {et~al.} 2013, The Astrophysical
  Journal, 764, 23, \dodoi{10.1088/0004-637x/764/1/23}

\bibitem[{{Sovers}(1991)}]{Sovers1991}
{Sovers}, O.~J. 1991, Observation model and parameter partials for the JPL VLBI
  parameter estimation software MODEST/1991, {JPL} publication 83-39 rev. 4,
  NASA

\bibitem[{Taktakishvili {et~al.}(2011)Taktakishvili, Pulkkinen, MacNeice,
  Kuznetsova, Hesse, \& Odstrcil}]{Taktakishvili2011}
Taktakishvili, A., Pulkkinen, A., MacNeice, P., {et~al.} 2011, Space Weather,
  9, \dodoi{10.1029/2010SW000642}

\bibitem[{Thompson(2006)}]{Thompson2006}
Thompson, W.~T. 2006, Astronomy {\&} Astrophysics, 449, 791,
  \dodoi{10.1051/0004-6361:20054262}

\bibitem[{Titov {et~al.}(2018)Titov, Girdiuk, Lambert, Lovell, McCallum,
  Shabala, McCallum, Mayer, Schartner, de~Witt, Shu, Melnikov, Ivanov,
  Mikhailov, Yi, Soja, Xia, \& Jiang}]{Titov2018}
Titov, O., Girdiuk, A., Lambert, S.~B., {et~al.} 2018, Astronomy {\&}
  Astrophysics, 618, A8, \dodoi{10.1051/0004-6361/201833459}

\bibitem[{T{\'{o}}th {et~al.}(2012)T{\'{o}}th, van~der Holst, Sokolov, Zeeuw,
  Gombosi, Fang, Manchester, Meng, Najib, Powell, Stout, Glocer, Ma, \&
  Opher}]{Toth2012}
T{\'{o}}th, G., van~der Holst, B., Sokolov, I.~V., {et~al.} 2012, Journal of
  Computational Physics, 231, 870, \dodoi{10.1016/j.jcp.2011.02.006}

\bibitem[{van~der Holst {et~al.}(2014)van~der Holst, Sokolov, Meng, Jin,
  W.~B.~Manchester, T{\'{o}}th, \& Gombosi}]{vanderholst2014}
van~der Holst, B., Sokolov, I.~V., Meng, X., {et~al.} 2014, The Astrophysical
  Journal, 782, 81, \dodoi{10.1088/0004-637x/782/2/81}

\end{thebibliography}

\end{document}